\documentclass[twocolumn,notitlepage,aps,prb,10pt,superscriptaddress,floatfix,letterpaper,reprint]{revtex4-1}	
\usepackage{hyperref}
\usepackage[usenames,dvipsnames]{color}
\usepackage{amsmath}
\usepackage[english]{babel}
\usepackage{amssymb}
\usepackage{graphicx}
\usepackage{epstopdf}
\usepackage[normalem]{ulem}
\usepackage{subfigure}
\usepackage{braket}
\usepackage{bbold}
\usepackage{placeins}
\usepackage[para]{threeparttable}
\usepackage{lipsum}
\usepackage{multirow}
\usepackage{bm}
\usepackage{siunitx}

\usepackage{todonotes}

\definecolor{linkcolor}{HTML}{399B03}
\definecolor{urlcolor}{HTML}{399B03}
\hypersetup{pdfstartview=FitH, linkcolor=linkcolor,urlcolor=urlcolor,colorlinks=true}
\hypersetup{
    unicode=false,          
    pdftoolbar=true,        
    pdfmenubar=true,        
    pdffitwindow=false,     
    pdfstartview={FitH},    
    pdfauthor={},         
    colorlinks=true,        
    linkcolor=blue,         
    citecolor=red,          
}

\begin{document}

\title{Error propagation in the fully self-consistent stochastic second-order Green's function method}
\author{Blair Winograd}
\affiliation{Department of Chemistry, University of Michigan, Ann Arbor, Michigan 48109, USA
}%
\author{Emanuel Gull}
\affiliation{Department of Physics, University of Michigan, Ann Arbor, Michigan 48109, USA
}%
\author{Dominika Zgid} 
\affiliation{Department of Chemistry, University of Michigan, Ann Arbor, Michigan 48109, USA
}%
\affiliation{Department of Physics, University of Michigan, Ann Arbor, Michigan 48109, USA
}%

\date{\today}
\begin{abstract}
We present an implementation of a fully self-consistent finite temperature second order Green's function perturbation theory (GF2) within the diagrammatic Monte Carlo framework. In contrast to the previous implementations of stochastic GF2 ({\it J. Chem. Phys.},{\bf 151}, 044144 (2019)), the current self-consistent stochastic GF2 does not introduce a systematic bias of the resulting electronic energies. Instead, the introduced implementation accounts for the stochastic errors appearing during the solution of the Dyson equation. We present an extensive discussion of the error handling necessary in a self-consistent procedure resulting in dressed Green's function lines. We test our method on a series of simple molecular examples.
\end{abstract}

\maketitle

\section{Introduction}\label{sec:Introduction}

While computationally inexpensive density functional theory (DFT) calculations are the workhorse of modern computational chemistry and materials science, there is a need for methods that perform calculations fully ab-initio in a systematically improvable manner. Such methods are necessary when the theoretical results cannot easily be checked against existing experiments and, consequently, a suitable DFT functional cannot be easily chosen to reproduce the experimental data. Moreover, with the advent of machine learning techniques and materials by design searches, the availability of computational ab-initio data is important to provide multiple unbiased calibration points. 

Many-body, finite temperature diagrammatic methods such as the self-consistent second order Green's function perturbation theory (GF2)~\cite{Dahlen05,Phillips14,Rusakov16,Iskakov19} or self-consistent GW~\cite{Hedin1965,Aryasetiawan98,Kutepov09,Kutepov20} are crucial to illustrate properties of many weakly correlated materials. For strongly correlated materials, these methods are frequently the first step in an embedding procedure \cite{DMFT_infinite_dim_Georges_Kotliar_1992,Georges96,Kotliar06,multitier_GW+DMFT_werner_2017,PhysRevLett.90.086402,doi:10.1021/acs.jpclett.8b01754,doi:10.1021/acs.jctc.8b00927,Tran_jcp_2015,Tran_jctc_2016,Tran_Shee_2017,Kananenka15,Zgid17,Lan17,SEET_NIO_MNO_2020}
that allows descriptions of systems that are too correlated to be evaluated quantitatively correctly within weakly correlated theories.

Finite temperature GF2 and finite temperature GW for molecular systems have a high computational cost of $\mathcal{O}(n^5n_{\tau})$ and $\mathcal{O}(n^6n_{\tau})$, respectively, if no additional approximations and simplifications are introduced. (We denote the number of orbitals as $n$ and the number of temperature grid points as $n_{\tau}$).
Consequently, at present, both of these methods can be used only for molecular systems of moderate size~\cite{Phillips14,PhysRevX.7.031059,PhysRevX.10.011041} 
and simple solids~\cite{Rusakov16,Iskakov19,Iskakov20}.

Stochastic sampling of the self-energy may make these methods cheaper, since many of the self-energy elements are small and may potentially be neglected.
This has led to the development of several stochastic methods for low order perturbation theory~\cite{mp2Rabani,dsGF2zgidbaer,sgf2takeshita,Dou20,ritake,sohiratasGF2,sohiratagf23}.
Most of theses methods either sample  the non-self-consistent version of the perturbation theory or do not change their sampling scheme in the presence of the  self-consistency. 

It is also possible to sample diagrammatic series stochastically to high orders.
Such methods, known as `Diagrammatic Monte Carlo', have been extensively used in lattice model systems~\cite{DiagQMC1998,PhysRevLett.81.2514,PhysRevB.77.020408,PhysRevB.77.125101,VANHOUCKE201095}. Applications to quantum chemistry Hamiltonians with many orbitals are still rare \cite{PhysRevX.7.031059,Li2020}, but recently an efficient method was introduced \cite{Li2020} that generalized these methods to quantum chemistry systems by sampling a bare (i.e. non-self-consistent) diagrammatic expansion up to high order. 

Bare series stand in contrast to self-consistent skeleton expansions such as the GF2 or GW, where an infinite number of diagrams is absorbed in renormalized propagators at the cost of introducing additional self-consistent equations. The stochastically sampled self-energy then enters non-linear equations that propagate stochastic errors and distort normally distributed data, necessitating a careful analysis of the resulting distributions. 

In this work, we present an algorithm for sampling the self-consistent finite temperature second order Green's function series, focusing on the analysis of errors and their propagation through the self-consistent equations.
The remainder of the paper is organized as follows. In Sec.~\ref{sec:Theory}, we introduce GF2 and discuss the sampling procedure. In Sec.~\ref{sec:StochasticError}, we discuss and analyze Monte Carlo errors and their propagation. In Sec.~\ref{sec:results}, we present results, and conclusions are shown in Sec.~\ref{sec:conclusion}.

\section{Theory}\label{sec:Theory}
\subsection{Description of the deterministic GF2 algorithm}
The GF2 method aims to compute the second-order self-energy $\Sigma(i\omega_n)$ in terms of the `renormalized' or `dressed' propagators $G(i\omega_n)$, where $\omega_n$ denotes the $n$th Matsubara frequency $\omega_n=\frac{(2n+1)\pi}{\beta}$, and $\beta=1/T$ is the inverse temperature.

The correlated Green's function is related to the self-energy by the Dyson equation
\begin{equation}
    G(i\omega_n)=[G_0(i\omega_n)^{-1}-\Sigma(i\omega_n)]^{-1},
\end{equation}
where $G_0$ is the Hartree--Fock Green's function defined as
\begin{equation}
G_0(i\omega_n)=[(i\omega_n + \mu)S-F]^{-1}.
\end{equation}
Here, $F$ and $S$ are the Fock and overlap matrices and $\mu$ is the chemical potential adjusted such that the correct number of particles is obtained.

\begin{figure}
      \includegraphics[width=\columnwidth]{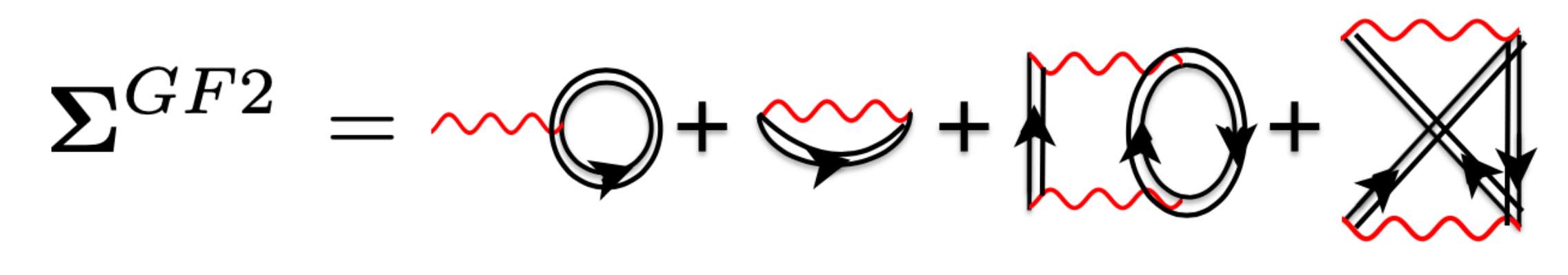}
  \caption{Diagrammatic representation of the self-consistent second order self-energy. Red lines: interaction terms $v$. Black lines: Green's functions $G$. The first two diagrams denote the frequency-independent contributions contained in the Fock matrix; the others represent the second-order correlation effects of Eq.~\ref{eq:sigma_second_order}.}
  \label{fig:selfenergydiag}
\end{figure}

The self-energy at second-order perturbation theory, $ \Sigma^{(2)}(\tau)$ that will be written simply as $ \Sigma(\tau)$ in the rest of this paper, is obtained from the Green's function as
\begin{align}\label{eq:sigma_second_order}
    \Sigma_{ij}(\tau)=-\sum_{klmnpq}&G_{kl}(\tau)G_{mn}(\tau)G_{pq}(-\tau) \nonumber \\ &\times v_{imqk}(2v_{lpnj}-v_{nplj}).
\end{align}
Here, $G(\tau)$ is the Green's function Fourier transformed from Matsubara frequency $\omega_n$ to imaginary time $\tau$, and $v$ are the two-body Coulomb integrals. An illustration of the first and second-order terms of the self-energy is given in Fig.~\ref{fig:selfenergydiag}. 

\begin{figure}
      \includegraphics[width=0.9\columnwidth]{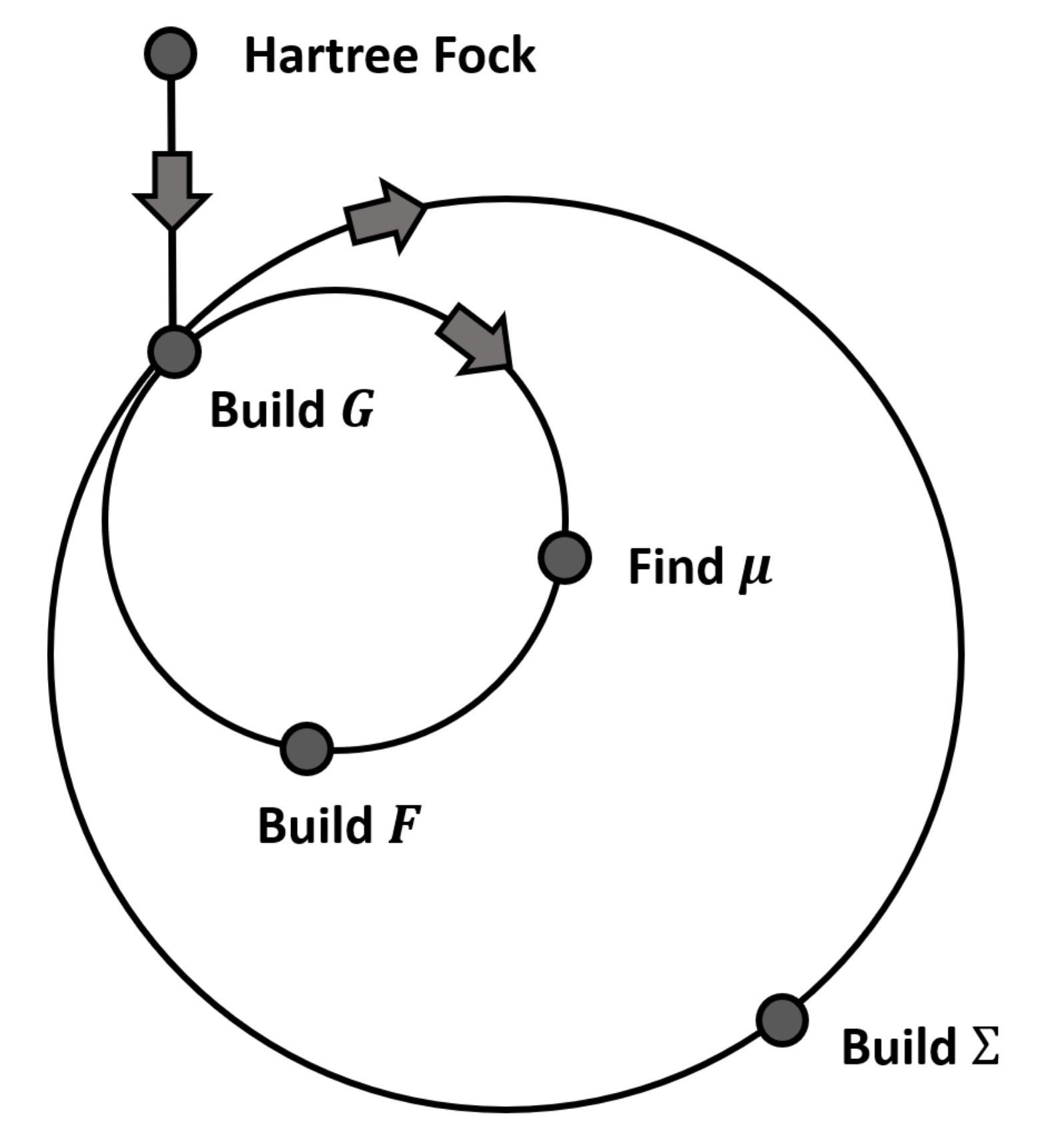}
  \caption{Iterative scheme for the self-consistent GF2 method.}
  \label{fig:birdeye_GF2}
\end{figure}

It is the evaluation of this second-order self-energy that dominates the computation of the GF2 equations.
A deterministic GF2 calculation involves the following steps:
\begin{enumerate}
    \item Run a Hartree-Fock (HF) or a DFT calculation for the system of interest.
    \item Build a starting Green's Function $G(i\omega_n)=[(i\omega_n +\mu)S-F]^{-1}$, with the Fock matrix from the earlier HF or DFT calculation.
    \item Fourier transform $G(i\omega_n)\to G(\tau)$.
    \item Using $G(\tau)$, build $\Sigma(\tau)$ according to Eq.~\ref{eq:sigma_second_order}.
    \item Fourier transform $\Sigma(\tau) \to \Sigma(i\omega_n)$.
    \item Rebuild the interacting Green's function $G(i\omega_n)=[(i\omega_n+ \mu)S-F-\Sigma(i\omega_n)]^{-1}$
    \item Compute the single-particle density matrix $\gamma=-2G(\tau=\beta)$ and adjust $\mu$ to obtain the desired electron number.
    \item Update the Fock matrix $F_{ij}=\sum_{kl}\gamma_{kl}(v_{ijkl}-0.5v_{ilkj})$ using the newly constructed density matrix.
    \item Evaluate the one- and two-body energy using the Galitskii -Migdal formula \cite{Galitzki58}.
    \item Rebuild the Green's function $G(i\omega_n)$  using the new Fock matrix.
    \item Pass this Green's function $G(i\omega_n)$ to step 3 of this cycle and  
    iterate until convergence in all quantities is reached.
\end{enumerate}
A sketch of this scheme is shown in Fig.~\ref{fig:birdeye_GF2}.

\subsection{Stochastic sampling of the second order self-energy}\label{sec:phifunc}
We now introduce an approach to sample the second-order self-energy discussed in the preceding section. For each $\Sigma_{ij}(\tau)$ in Eq.~\ref{eq:sigma_second_order} (with $i$ and $j$ denoting orbital indices), a sampling over six discrete internal orbital indices ($k$, $l$, $m$, $n$, $p$, and $q$) is required. However, many elements of $\Sigma_{ij}(\tau)$ are small or zero, making a direct sampling of each of the indices inefficient. Instead, we attempt to perform a sampling that generates the internal indices as well as the external indices $i,$ $j$, and $\tau$, sampling at once over $8$ orbital indices and time.

It is advantageous to compress and truncate the Green's functions and interaction vertices as much as possible before any sampling is attempted. For each time point, the Green's function $G_{ij}(\tau)=G_{ji}(\tau)$ is a symmetric matrix and can therefore be decomposed by an eigen\textcolor{red}{value} decomposition in orbital space into
\begin{align}
    G_{ij}(\tau)=G_i^\lambda(\tau) D^\lambda(\tau) G_j^\lambda(\tau). 
\end{align}
Here,  $D^\lambda$ is the diagonal matrix of eigenvalues. The expansion can then be truncated for small $D^\lambda$.

The two-electron integrals $v$ can similarly be decomposed using either eigenvalue or Cholesky decomposition into
\begin{align}
v_{ijkl}=v_{ij}^\alpha D^{\alpha}v_{kl}^\alpha,
\end{align}
and truncated in the index $\alpha$ for small $D^\alpha$ \cite{Schrader62,Baerends73,Aquilante11}. Combining contractions over the orbital indices $i$ and $j$ in $v_{ij}^\alpha$ with the decomposed Green's functions $G_i^\lambda$ and $G_j^\mu$ allows us to define the compact object
\begin{equation}
x_{\lambda\mu}^\alpha(\tau) = G_i^\lambda(\tau) v_{ij}^\alpha G_j^\mu(-\tau).
\end{equation}

Naively generating Monte Carlo configurations of these indices and times is inefficient, as the diagrams with large contribution to the second order self-energy are not uniformly distributed over all indices. Rather, we employ an importance sampling procedure that aims to focus the effort in parts of configuration space where large contributions are expected. This procedure requires a weight function on the space of diagrams, which is large predominantly where diagrams are large. 

\begin{figure}
      \includegraphics[width=0.9\columnwidth]{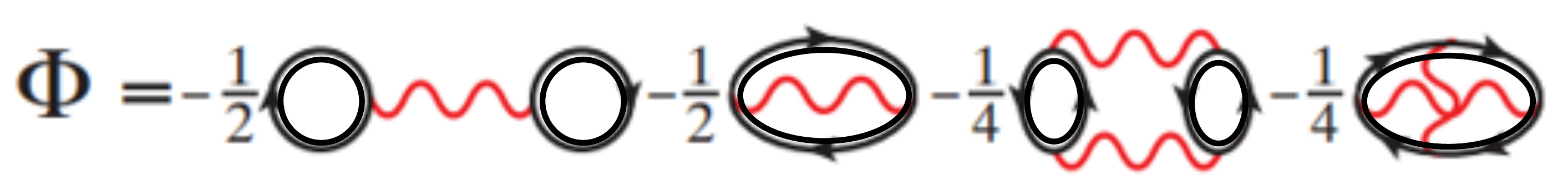}
  \caption{Diagrams of the $\Phi$ functional at second order.}
  \label{fig:phidiagrams}
\end{figure}

In this work, we choose to use terms of the Luttinger--Ward functional \cite{Luttinger60},
\begin{align}
    \Phi=\sum^{\infty}_{m=1}\frac{1}{2m}{\rm Tr}[\sum_n\Sigma^{(m)}(i\omega_n)G(i\omega_n)],
\end{align}
which is related to the self-energy as
$\frac{\delta\Phi}{\delta G_{ji}(\tau)}=\Sigma_{ij}(\tau),$
as an inspiration for the sampling weights. The functional is closely related to the free energy \cite{Luttinger60}. Diagrammatically, it contains the sum of all closed two-particle irreducible skeleton diagrams, as illustrated in Fig.~\ref{fig:phidiagrams}.

The explicit expression at second-order is
\begin{align}
\label{eqn:phisecondorder2}
\Phi^{(2)} &=\frac{1}{4}\sum_{ijklmnpq}\int d\tau v_{imqk}(2v_{jnlp}-v_{jlnp}) \nonumber \\ &\times G_{ij}(-\tau)G_{kl}(\tau)G_{mn}(\tau)G_{pq}(-\tau).
\end{align}
In terms of the contracted expressions for $v$ and $G$, the functional $\Phi$ at second order becomes
\begin{align}
\Phi^{(2)} &= \frac{1}{4}\int d\tau \sum_{\alpha \beta}^{n^v_d}\sum_{\lambda \mu \nu \sigma}^{n^g_d} x_{\lambda \mu}^\alpha(\tau)x_{\sigma \nu}^\alpha(\tau)D^\alpha D^\beta \nonumber \\ 
&\times D^\lambda(-\tau)D^\mu(\tau)D^\nu(\tau)D^\sigma(-\tau) \nonumber \\ &\times (2x_{\lambda \mu}^\beta(\tau)x_{\sigma \nu}^\beta(\tau)-x_{\lambda \nu}^\beta(\tau)x_{\sigma \mu}^\beta(\tau)), 
\label{eq:phifunctionaldecomposed}
\end{align}
where $n^v_d$ and $n^g_d$ denote the number of vertex and Green's function decomposition indices kept. Due to the decay of the eigenvalues in the decomposition, these values can often be chosen much smaller than the indices in the original problem. 

Discretizing the time integral on a non-uniform imaginary time grid (here we used a power grid \cite{Ku2000} but other grids are possible \cite{Boehnke2011,Gull2018,PhysRevB.96.035147,Hugo_grid_2020}) approximates the integral in Eq.~\ref{eq:phifunctionaldecomposed} as the sum
\begin{align}
\Phi^{(2)} &= \frac{1}{4}\sum_\tau^{n_\tau}\sum_{\alpha \beta}^{n^v_d}\sum_{\lambda \mu \nu \sigma}^{n^g_d} x_{\lambda \mu}^\alpha(\tau)x_{\sigma \nu}^\alpha(\tau) \nonumber \\ &\times D^\alpha D^\beta 
D^\lambda(-\tau)D^\mu(\tau)D^\nu(\tau)D^\sigma(-\tau) \nonumber \\ &\times (2x_{\lambda \mu}^\beta(\tau)x_{\sigma \nu}^\beta(\tau)-x_{\lambda \nu}^\beta(\tau)x_{\sigma \mu}^\beta(\tau))w_\tau, \label{eq:phidiscretesum} \nonumber \\
&=\sum_\tau^{n_\tau}\sum_{\alpha \beta}^{n^v_d}\sum_{\lambda \mu \nu \sigma}^{n^g_d} \phi(\lambda\mu\nu\sigma;\alpha\beta;\tau),
\end{align}
where $w_\tau$ denotes the integration weight with $\sum_\tau w_\tau = \beta$. 

While  $\Phi^{(2)}$ in Eq.~\ref{eq:phidiscretesum} is  positive and real, individual terms in this sum may have any sign or complex phase. In order to make a sampling possible, we define the weight function
\begin{align}
    p(\lambda\mu\nu\sigma;\alpha\beta;\tau)=|\phi(\lambda\mu\nu\sigma;\alpha\beta;\tau)|.
\end{align}
This weight function is positive, normalizable, and hence allows an importance sampling procedure of configurations $c=\{\lambda\mu\nu\sigma;\alpha\beta;\tau\}$ with weight $p(c)$.

To obtain the overall magnitude of the quantities measured, we explicitly compute $\Phi^{(2)}$ for a small subset of typically 1-10 orbitals or just the diagonal contributions, and then use the overlap of the sampled second order contribution with the explicitly computed subset to normalize all expressions.

\subsection{Sampling Scheme}\label{sec:metropolis}
\begin{figure*}[bth]
\includegraphics[width=0.3\textwidth]{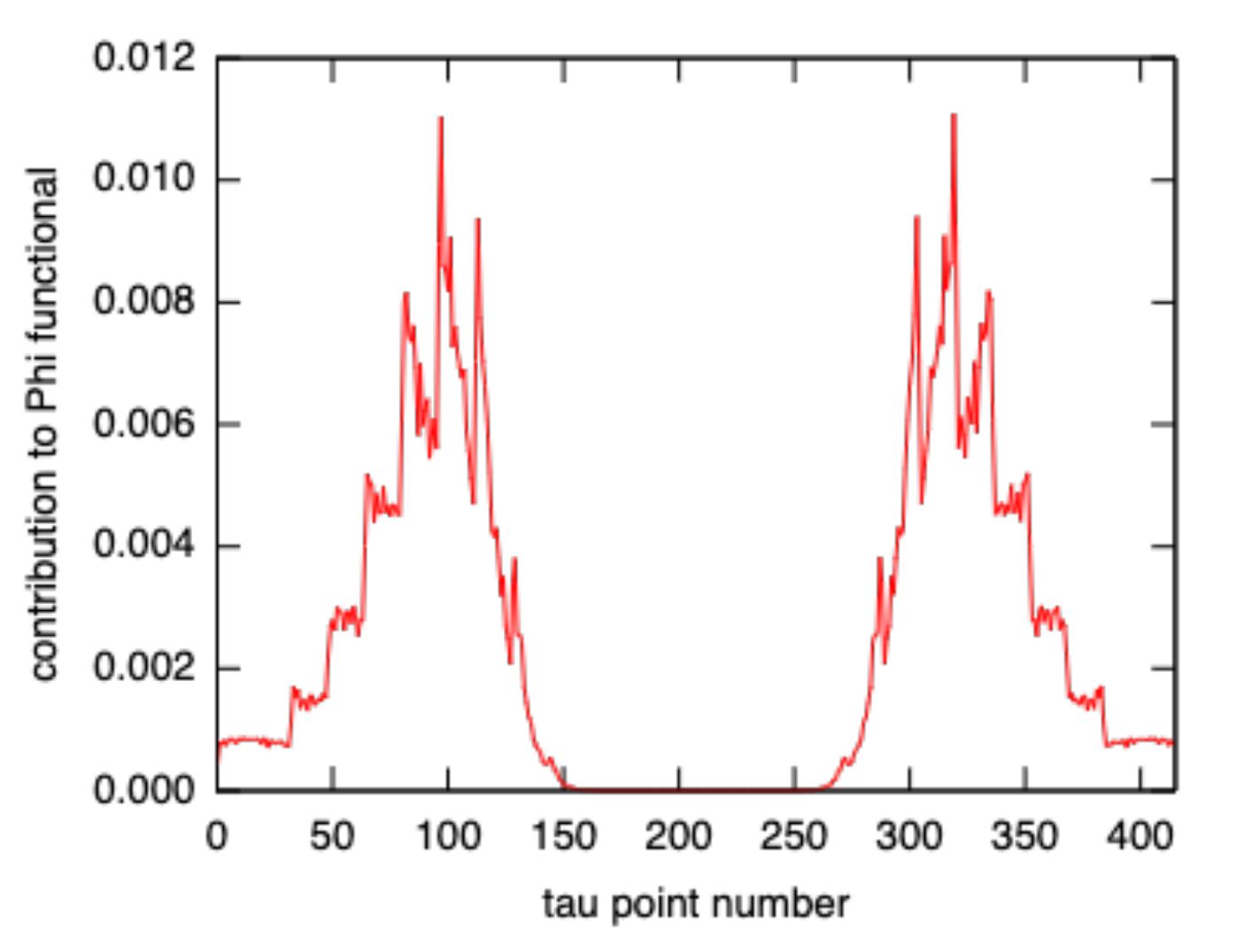}
\includegraphics[width=0.3\textwidth]{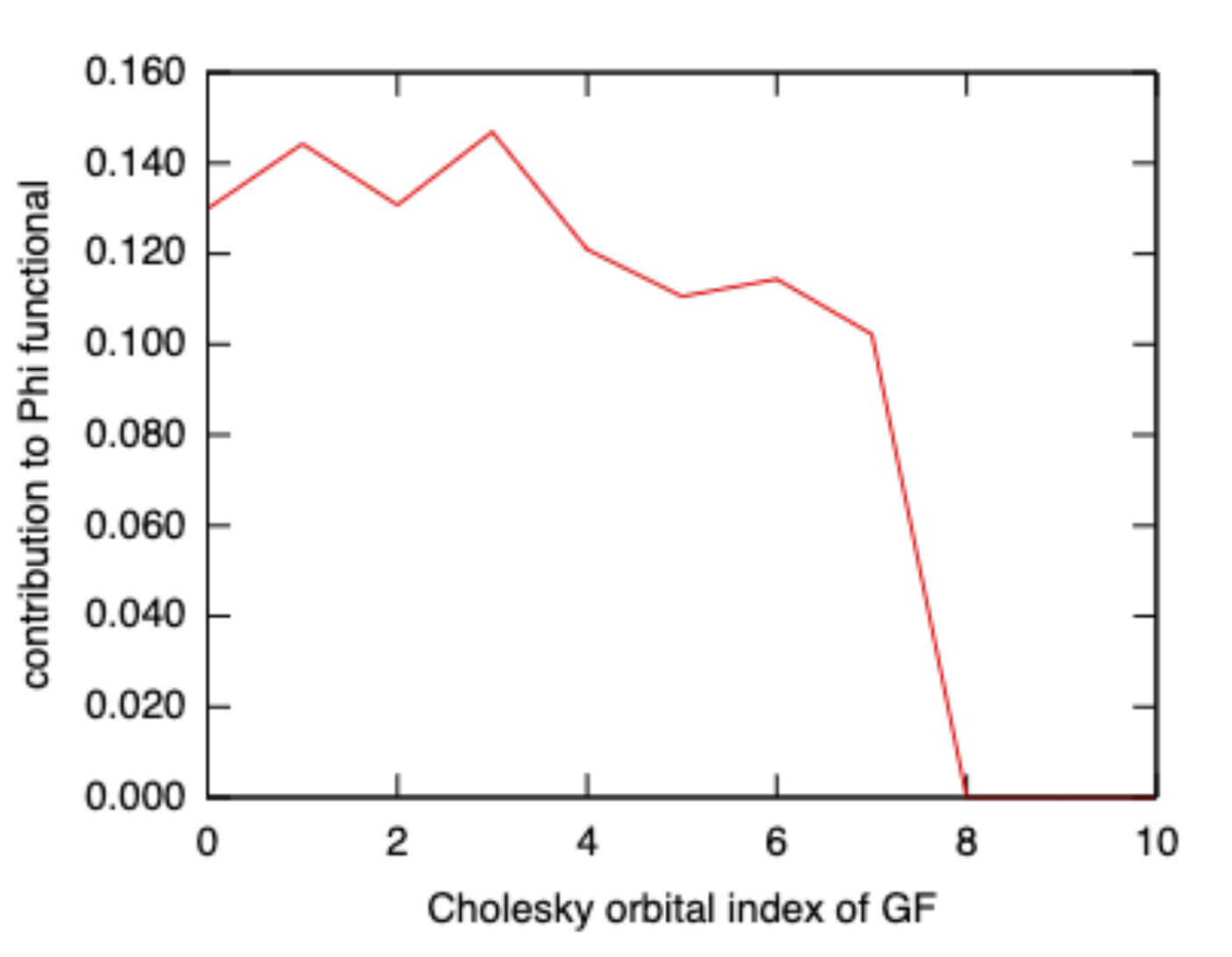}
\includegraphics[width=0.3\textwidth]{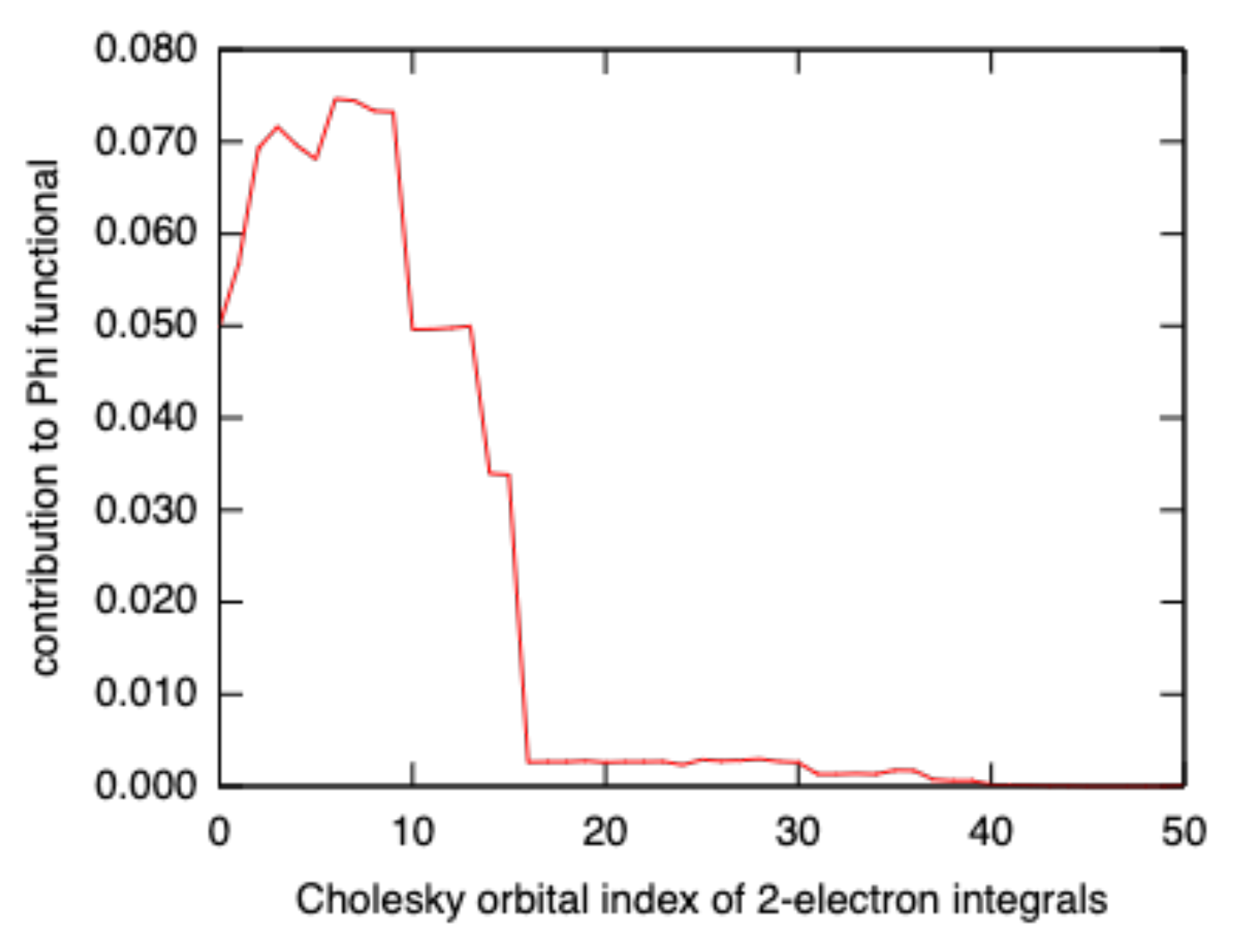}
\caption{Contributions to $\Phi^{(2)}$. Left panel: Histogram of configurations visited in $\tau$ (the underlying power law discretization of the imaginary time grid is visible as successive steps in the data). Middle panel: histogram of configurations in the Green's functions. Right panel: histogram of configurations in the vertex indices. Data obtained for a chain of 10 hydrogen atoms in a minimal STO-3G basis set, R=1 {\AA}.} 
  \label{fig:sampling}
\end{figure*}

We sample configurations with weight $p$ in the configuration space with three distinct types of updates: the change of a time slice $\tau$, the change of one of the four Green's function orbital indices $\lambda \mu \nu \sigma$, and the change of the interaction indexes $\alpha \beta$. Configurations are proposed, accepted, or rejected according to a Metropolis sampling scheme. Histograms for the configurations visited are shown in Fig.~\ref{fig:sampling}. With these three update types, any possible configuration can be generated from any other term in a finite number of steps.
Updating an orbital index only changes one Green’s function 
and one interaction vertex line, while updating a $\tau$ point changes all Green’s function lines.

\subsection{Measuring the self-energy}\label{sec:measuring self-energy}
During the sampling of configurations of $\Phi^{(2)}$, contributions to the second-order self-energy can be obtained by `cutting a Green's function line', $\frac{\delta \Phi}{\delta G_{ij}}=\Sigma_{ji}$. 
Given a configuration with Green's function, vertex, and time indices, we measure the following four contributions to the self-energy:
\begin{eqnarray}
     \Sigma_{lk}(-\tau_r)=& \frac{w(c)}{G_{kl}(\tau_r)w_r},\\
     \Sigma_{nm}(-\tau_r)=&  \frac{w(c)}{G_{mn}(\tau_r)w_r},\\
      \Sigma_{qp}(\tau_r)=& \frac{w(c)}{G_{pq}(-\tau_r)w_r},\\
      \Sigma_{ji}(\tau_r)=& \frac{w(c)}{G_{ij}(-\tau_r)w_r}.
\end{eqnarray}
Since the self-energy fulfills the relationship $\Sigma_{ij}(-\tau_r) = -\Sigma_{ji}(\beta-\tau_r)$, the first two expression result in
\begin{eqnarray}
     \Sigma_{kl}(\beta-\tau_r)=& \frac{w(c)}{G_{kl}(\tau_r)w_r},\\
     \Sigma_{mn}(\beta-\tau_r)=&  \frac{w(c)}{G_{mn}(\tau_r)w_r}.
\end{eqnarray}

\section{Error Propagation in the Dyson Equation}\label{sec:StochasticError}
Any self-consistent Green's function method that obtains samples for the self-energy, which are then used to obtain a Green's function for the next iteration, will need to account for the error propagation of the self-energy contributions through the Dyson equation $G(i\omega_n)=[(\mu + i\omega_n)S-F-\Sigma(i\omega_n)]^{-1}$. This equation is non-linear and therefore may amplify or skew the normal distribution of the Monte Carlo samples of the self-energy.

A careful analysis of this error propagation needs to be performed in order to obtain unbiased self-consistent solutions from stochastic iterative methods. Several popular methods are available, including the jackknife and the bootstrap method~\cite{Efron82,Efron94}. Here we employ the jackknife resampling method which, given a set of $n$ independent estimates $\Sigma_j$ of the self-enery $\Sigma$, provides estimates for the  mean and error of the Green's function while properly accounting for the non-linear error propagation by the Dyson equation. The method starts by performing the average of all estimates except for the $i^{th}$ one,
\begin{align}
    {\bar {\Sigma}}_{i}=\frac{1}{n-1}\sum^n_{j=1,j\ne i} \Sigma_j,
\end{align}
for  $i=1,\dots,n$. The Green's function, $G(\Sigma)$, resulting from the Dyson equation  is then evaluated for each ${\bar {\Sigma}}_{i}$, resulting in the values $G_i =G({\bar {\Sigma}}_{i})$ for the Green's function, as well as for the average of all values, $G_0=G(\frac{1}{n}\sum_i \Sigma_i)$. The jackknife estimate for $G$ will then be
\begin{align}
  G=G_0 - (n-1)(\bar{G}-G_0),
\end{align}
with statistical error
\begin{align}
  \Delta G=\sqrt{n-1}\left[\frac{1}{n}\sum_i G_i^2-{\bar G}^2\right]^{1/2},
\end{align}
where
\begin{align}
   \bar G=\frac{1}{n}\sum_i G_i.
\end{align}
The term $(n-1)(\bar{G}-G_0)$ is in general non-zero in non-linear transformations. If it is omitted, the means may show a systematic shift (often larger than the naively estimated error bars) that will only disappear as the number of samples is increased.

\begin{figure}
      \includegraphics[width=\columnwidth]{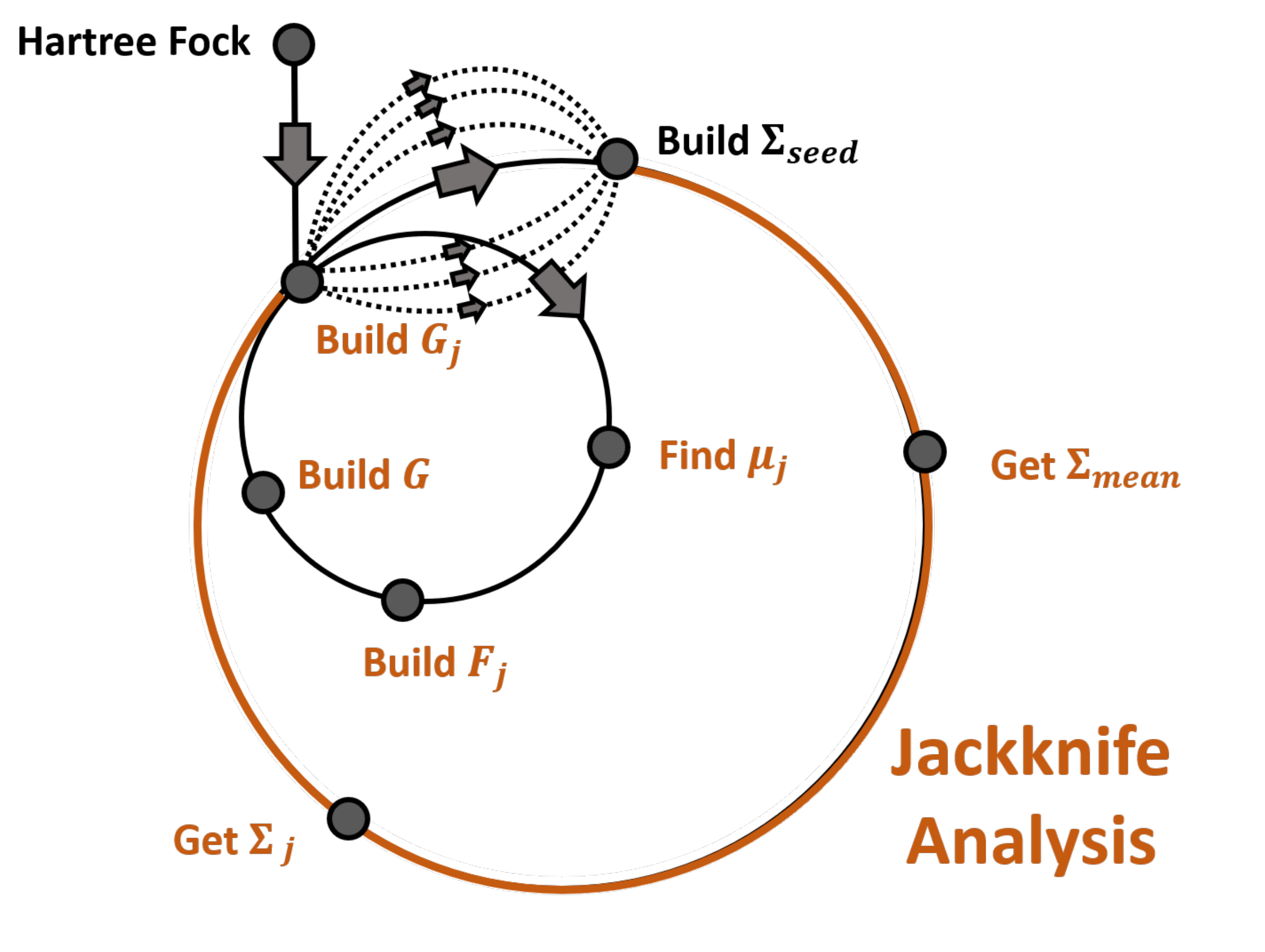}
  \caption{Modified scheme of  Fig.~\ref{fig:birdeye_GF2} accounting for non-linear error propagation} \label{fig:jackknife_scheme}
\end{figure}

In summary, the GF2 algorithm with jackknife analysis, illustrated in Fig.~\ref{fig:jackknife_scheme}, is as follows:
\begin{enumerate}
    \item Run either a HF or a DFT  calculation for the system of interest, obtain the Fock matrix.
    \item Build a starting Green's Function $G(i\omega_n)=[(i\omega_n +\mu)S-F]^{-1}$, with the Fock matrix from the HF or DFT calculation.
    \item Fourier transform $G(i\omega_n)\to G(\tau)$.
    \item Starting from a set of different random seeds, sample the second order contribution to the $\Phi$-functional as described above and measure the self-energy $\Sigma(\tau)$. 
    \item Collect all estimates and compute $\bar{\Sigma}$, $\bar{\Sigma}_i$. We call these quantities `subestimates'.
    \item Fourier transform the subestimates to time and compute the correlated Green's function $[G(i\omega_n)_i]=[(i\omega_n+\mu)S-F-\Bar{\Sigma}_i(i\omega_n)]^{-1}$ for each $i$.
    \item Update the chemical potential, $\mu$, and build correlated single-particle density matrix with correct particle number for each subestimate.
    \item Update the Fock matrix for each subestimate.
    \item Using subestimates of Green's function and self-energy evaluate both the one- and two-body electronic energies and their errors.
    \item Evaluate the correlated Green's function $G(i\omega_n)$ and Fock matrix, compute their errors, and pass them to point 3. Iterate until  the jackknife estimates of both the one-body and the two-body energy are converged within error bars. 
\end{enumerate}

\section{Results}\label{sec:results}
In this section, we will provide a calibration of
the one-body, two-body, and subsequent total energies for a few simple systems. The values determined from the first iteration of stochastic GF2 and the fully self-consistent GF2 procedure will be compared to those from the deterministic procedure. Our implementation is built on a version of the ALPS libraries \cite{ALPS}.  The Cholesky decompositions  for all two-electron integrals are obtained by Dalton \cite{Aidas14}.

\subsection{Error Following One Iteration of stochastic GF2}
In the first iteration of the stochastic GF2 scheme, the Green's functions and Fock matrices used to calculate the self-energy are identical to those in the first iteration of deterministic GF2.  This simplifies the analysis of errors in the Monte Carlo procedure. 

For a correctly estimated error, two conditions should be met. First, the resulting one-body and two-body deterministic energies should typically fall within two error bars of the stochastic one-body and two-body energies. Second, as the number of samples considered increase, the error bars on the resulting stochastic energies should systematically decrease as $1/\sqrt{N}$, where $N$ is the number of samples, and the values converge to the deterministically computed result. 

The following results show the validation of error control for two toy systems. 

\begin{figure*}[tbh]
\begin{center}
      \includegraphics[width=0.95\columnwidth]{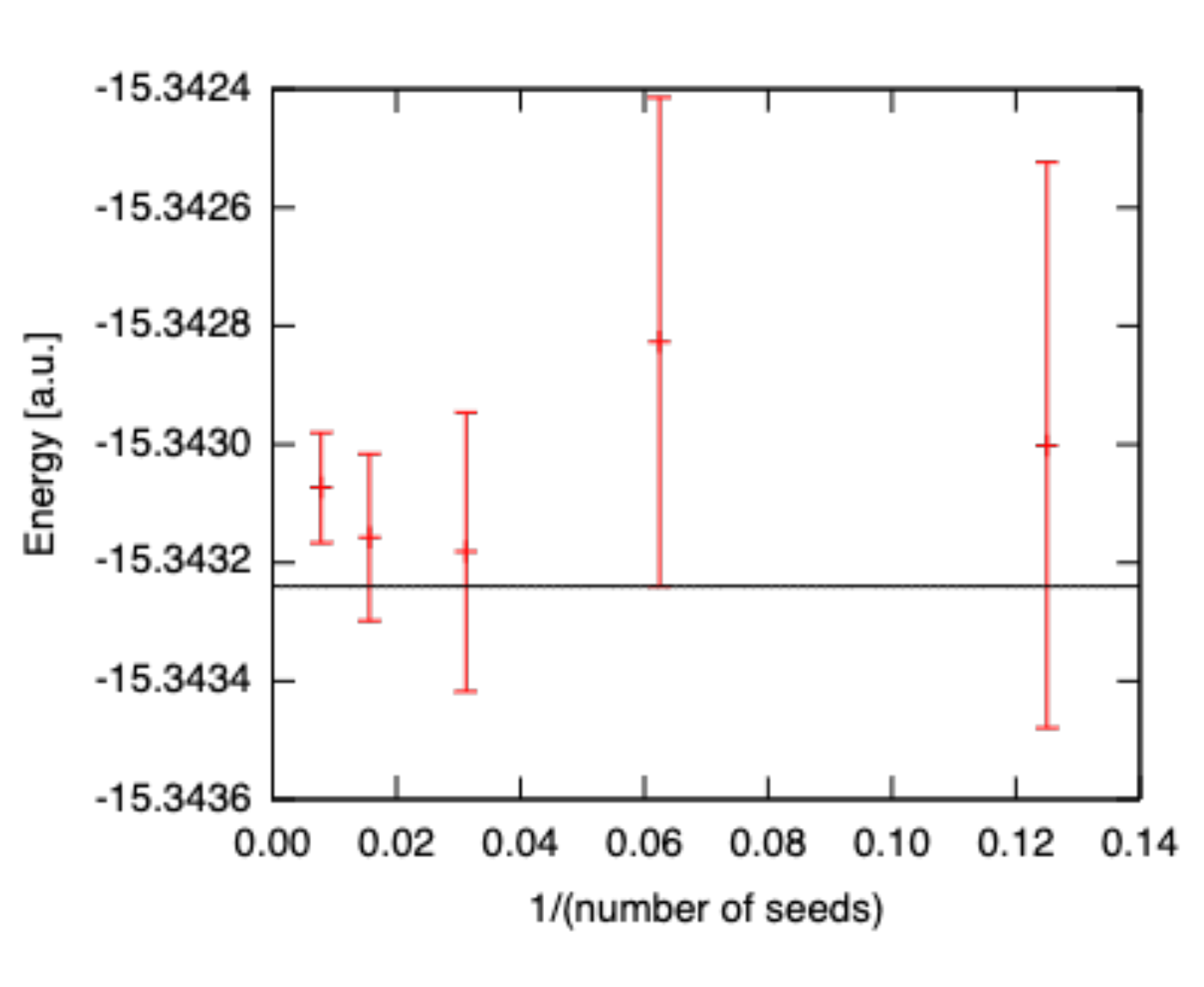}
      \includegraphics[width=0.95\columnwidth]{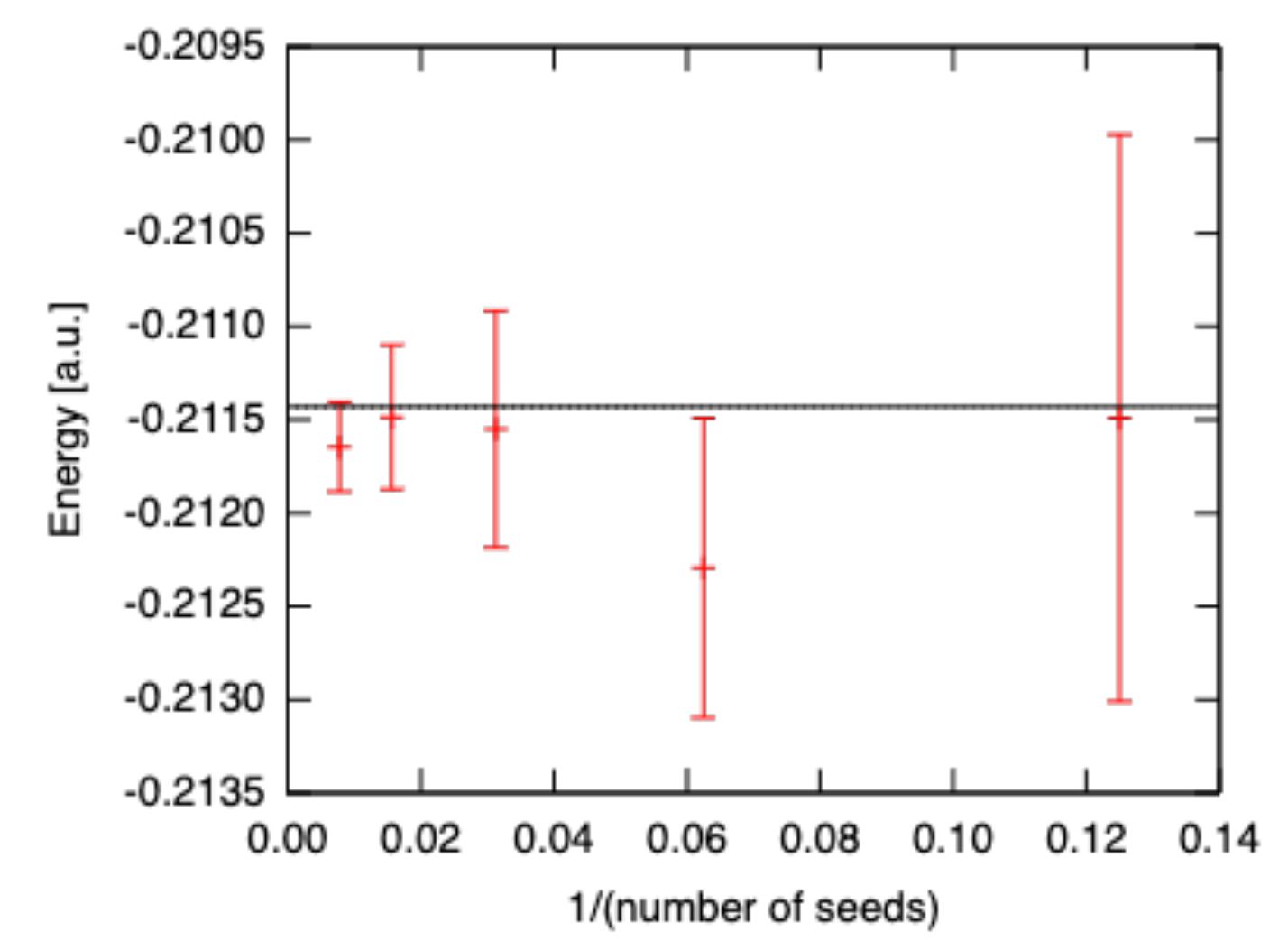}
\end{center}
\caption{One-body (left panel) and two-body (right panel) energy of a hydrogen chain H$_{10}$ in a STO-3G basis at distance of $1$ {\AA}, plotted versus the inverse of the number of Monte Carlo steps taken. Black line: reference energy evaluated deterministically. Error bars denote one-$\sigma$ errors.} 
  \label{fig:plot_ch10_1b2b}
\end{figure*}

First, we choose a system of 10 hydrogens placed one {\AA} apart.  A minimal STO-3G basis set is used.  The normalization subspace uses $n_t^v=1$. 
 We perform independent Monte Carlo simulations (starting from different seeds), each having a fixed number of $10^7$ Monte Carlo sampling steps. We then average results from $8$, $16$, $32$, $64$ and $128$ seeds using the jackknife procedure outlined above to obtain one-body and two-body energies. The results shown in Fig.~\ref{fig:plot_ch10_1b2b} illustrate that as the number of samples is increased, the stochastic result converges to the deterministic result with error bars that are consistent with expectations.
  
\begin{figure*}[bth]
\begin{center}
\includegraphics[width=\columnwidth]{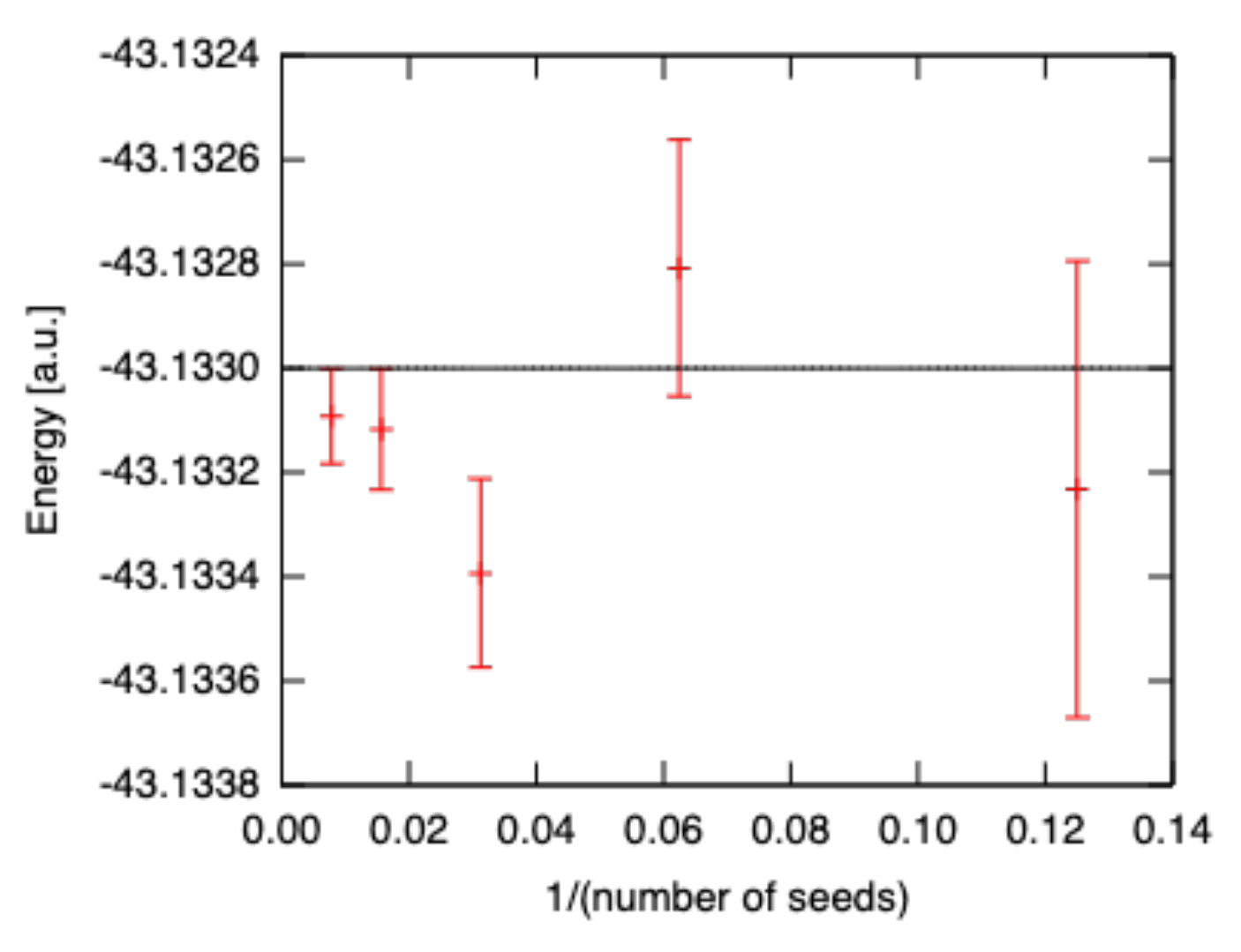}
\includegraphics[width=\columnwidth]{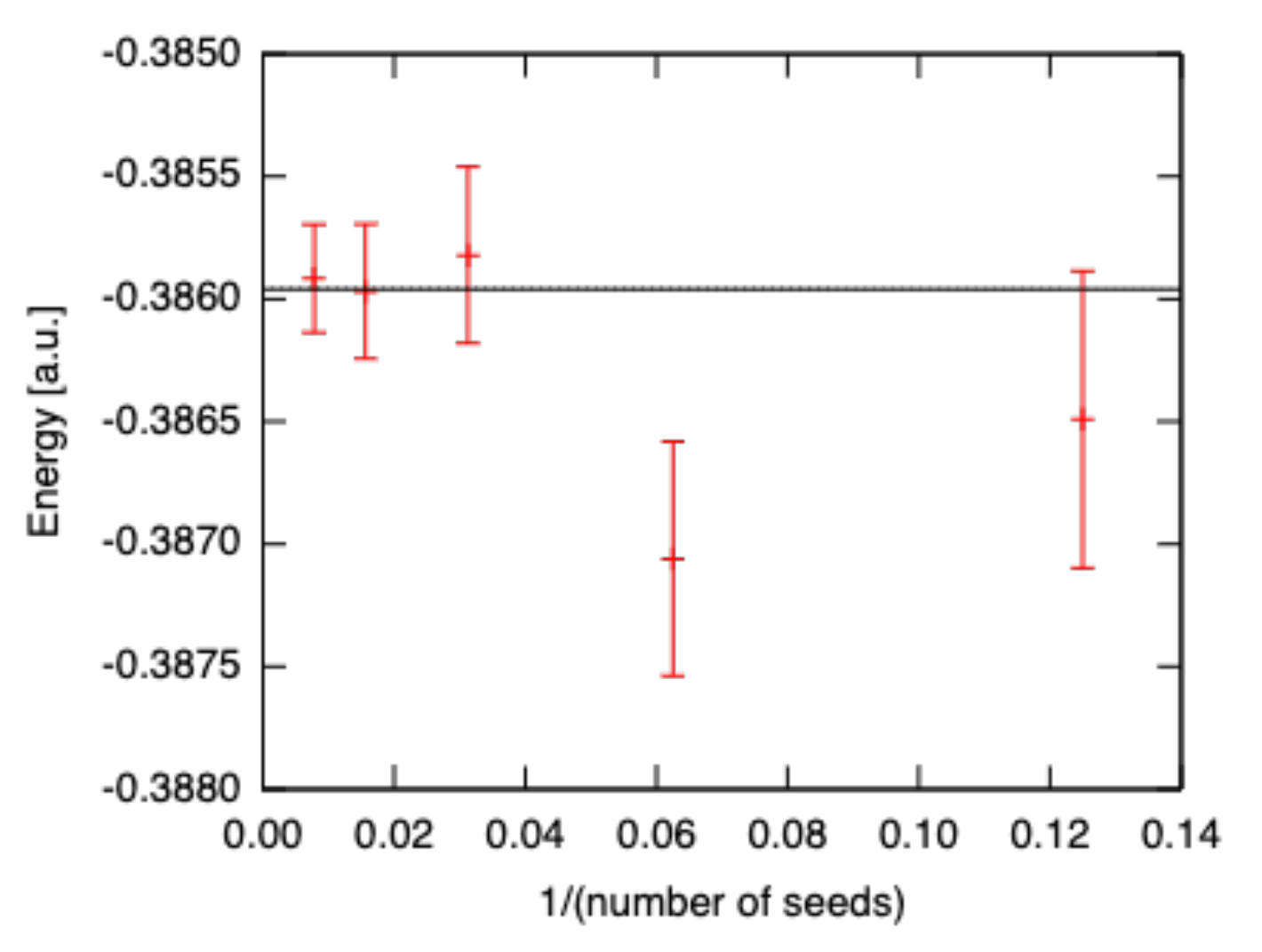}
\end{center}
\caption{One-body (left panel) and two-body (right panel) energy of a $4\times 4$ hydrogen square lattice in a STO-3G basis at distance of 1 {\AA}, plotted versus the inverse of the number of Monte Carlo steps taken. Black line: reference energy evaluated deterministically. Error bars denote one-$\sigma$ errors.} \label{fig:plot_pl44_1b2b}
\end{figure*}

Second, we chose a system of 16 hydrogens spaced $1$ {\AA} apart and organized into a 4$\times$4 square lattice in a minimal STO-3G basis.  The normalization subspace is evaluated using $n_t^v=3$. We follow the setup described above and obtain jackknifed results using one-body and two-body energies for $8$, $16$, $32$, $64$ and $128$ seeds, see Fig.~\ref{fig:plot_pl44_1b2b}.
As in the case of a linear chain, for the $4\times 4$ hydrogen square lattice the stochastic errors converge as expected. We emphasize that naive evaluation of the stochastic errors and means that does not account for the non-linearities of the energy evaluation would show an underestimation of the errors and a systematic bias in the stochastic mean, outside of error bars. Neglecting a proper error propagation therefore may lead to wrong conclusions.

\subsection{Error  of fully self-consistent GF2}
While in the first iteration of the stochastic GF2 procedure only a single inversion of the Green's function using Dyson equation is performed. When the stochastic GF2 procedure is executed fully self-consistently, a jackknife averaged Green's function and Fock matrix obtained at iteration $n$ enter the input of iteration $n+1$. The errors on these input quantities may propagate through repeated solutions of the GF2 equations and, potentially, amplify. 
If this error is controlled correctly, the deterministically evaluated one-body and two-body energies should be consistent with the means and error bars of the stochastic procedure. As has been reported in Ref.~\cite{dsGF2zgidbaer}, neglecting the error propagation in the self-consistent equations leads to a systematic bias in the results.

\begin{figure}
     \includegraphics[width=\columnwidth]{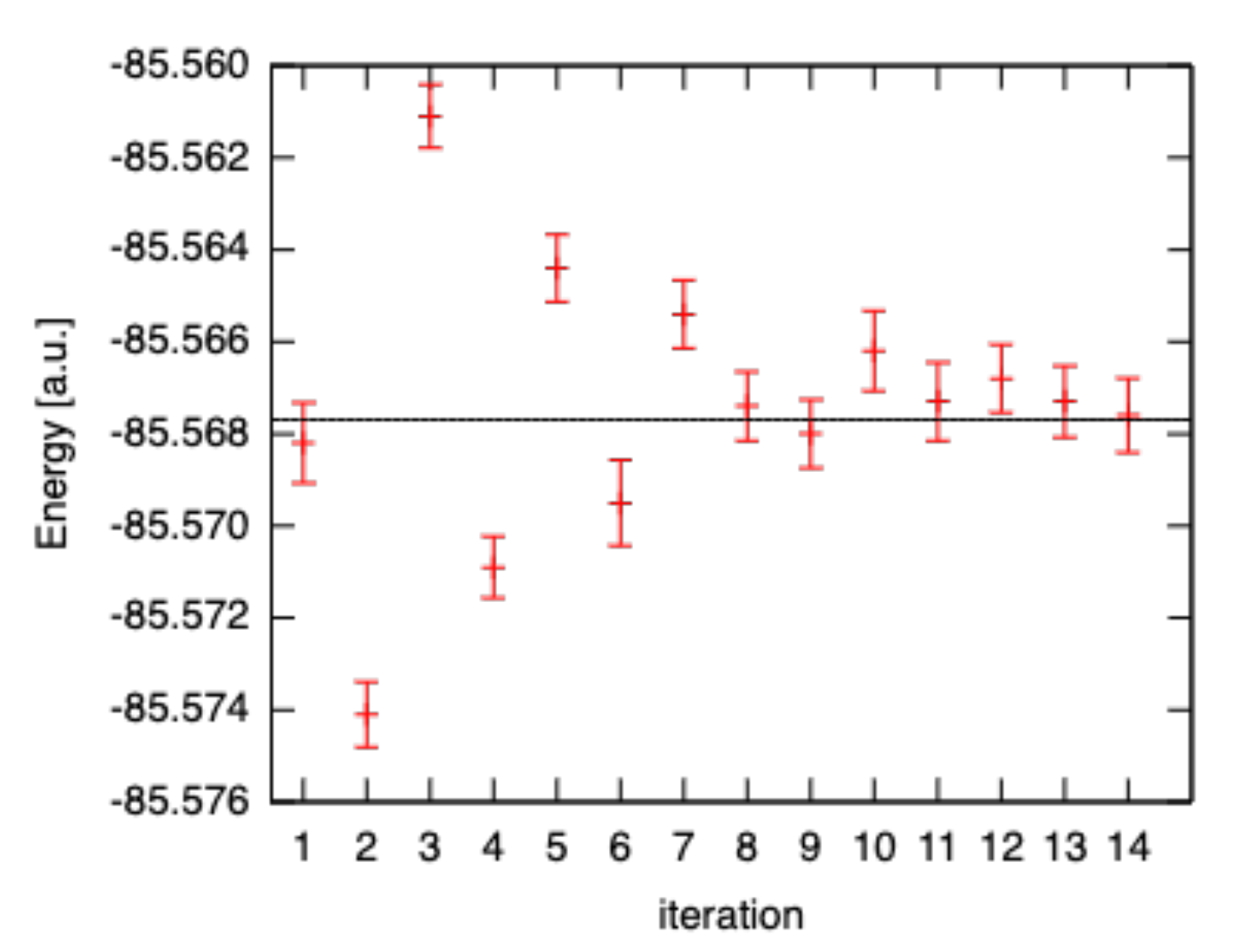}
  \caption{Total energy with error bars (red) for a water molecule in the cc-pVDZ basis as evaluated in the stochastic GF2 procedure, plotted as a function of iterations.  Black horizontal line: deterministic GF2 result for the converged total energy. Overall, the procedure converges within 14 iterations.}
  \label{fig:water8H2O_scf}
\end{figure}

To study the fully self-consistent scheme and consider the error between each iteration until convergence is reached, we chose to perform stochastic GF2 on a water molecule.
We use the cc-pVDZ basis set for both the deterministic and stochastic GF2 calculations.  We perform $10^8$ Monte Carlo sampling steps per run, and combine results from $128$ independent runs in a jackknife analysis. The normalization subspace is truncated at $n_t^v=6$.  
In Fig.~\ref{fig:water8H2O_scf}, the jackknife results on the total energy are given for each iteration of the calculation.  Each iteration is evaluated with $128$ seeds. 

\begin{figure*}[bt]
\begin{center}
     \includegraphics[width=0.32\textwidth]{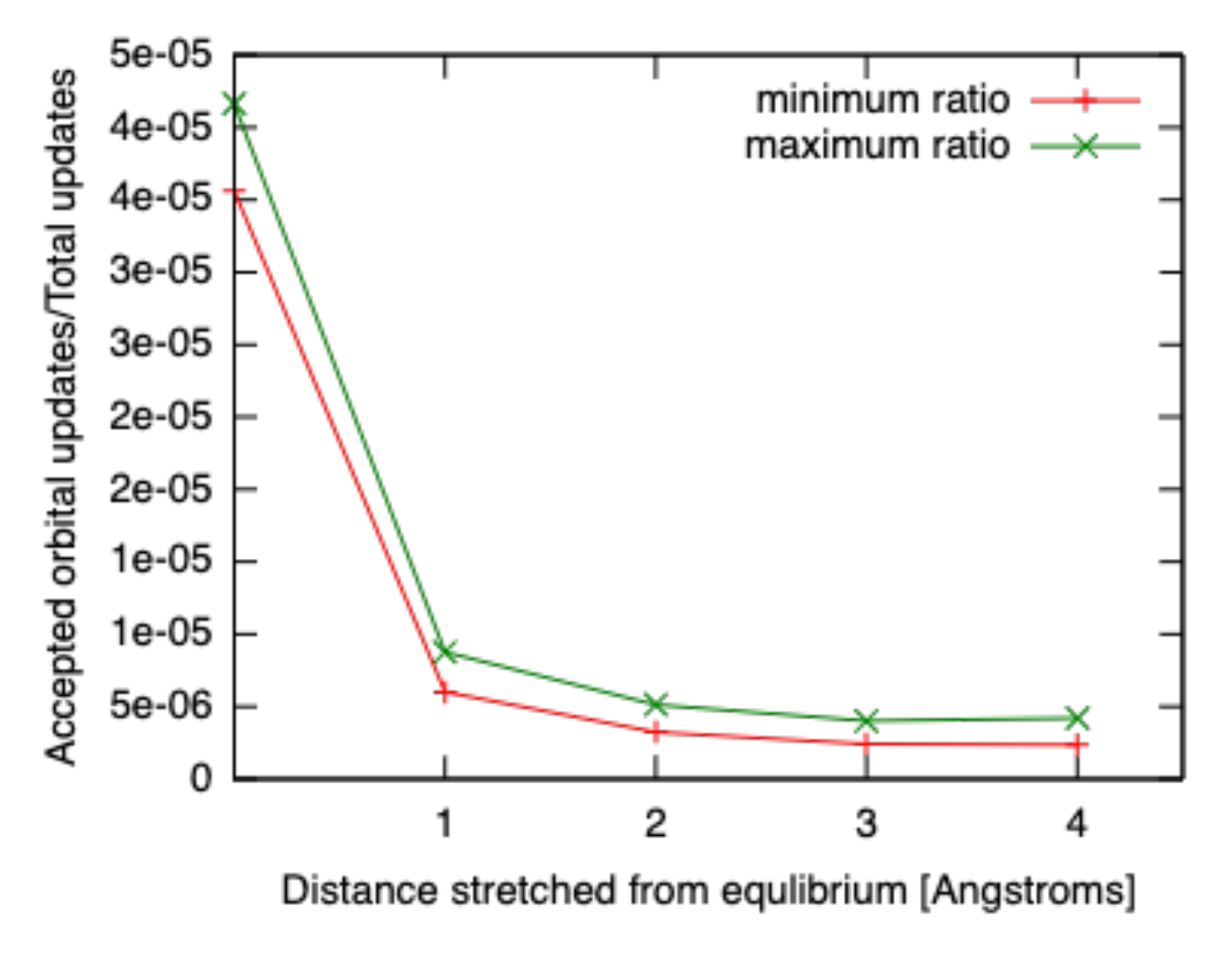}
     \includegraphics[width=0.32\textwidth]{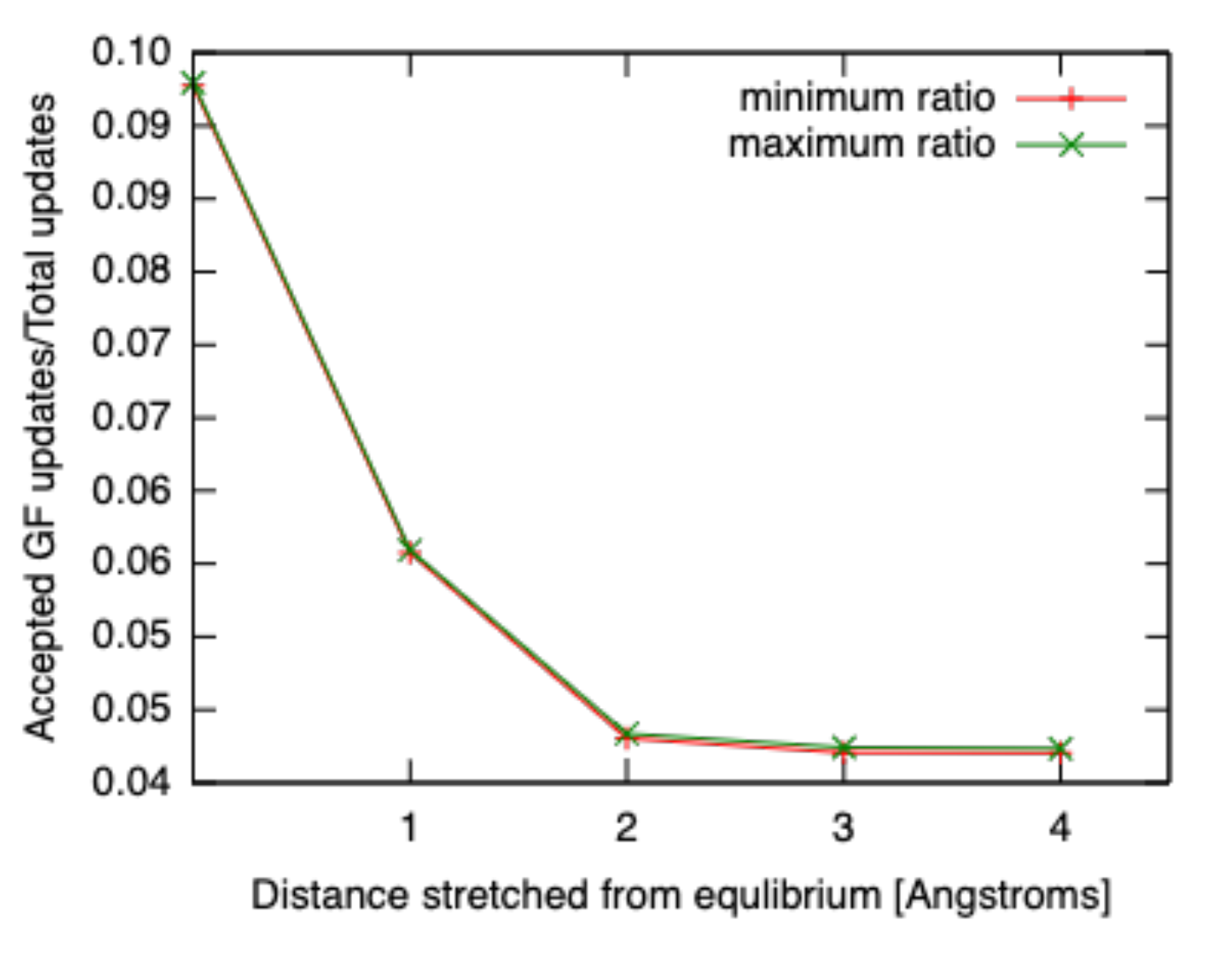}
     \includegraphics[width=0.32\textwidth]{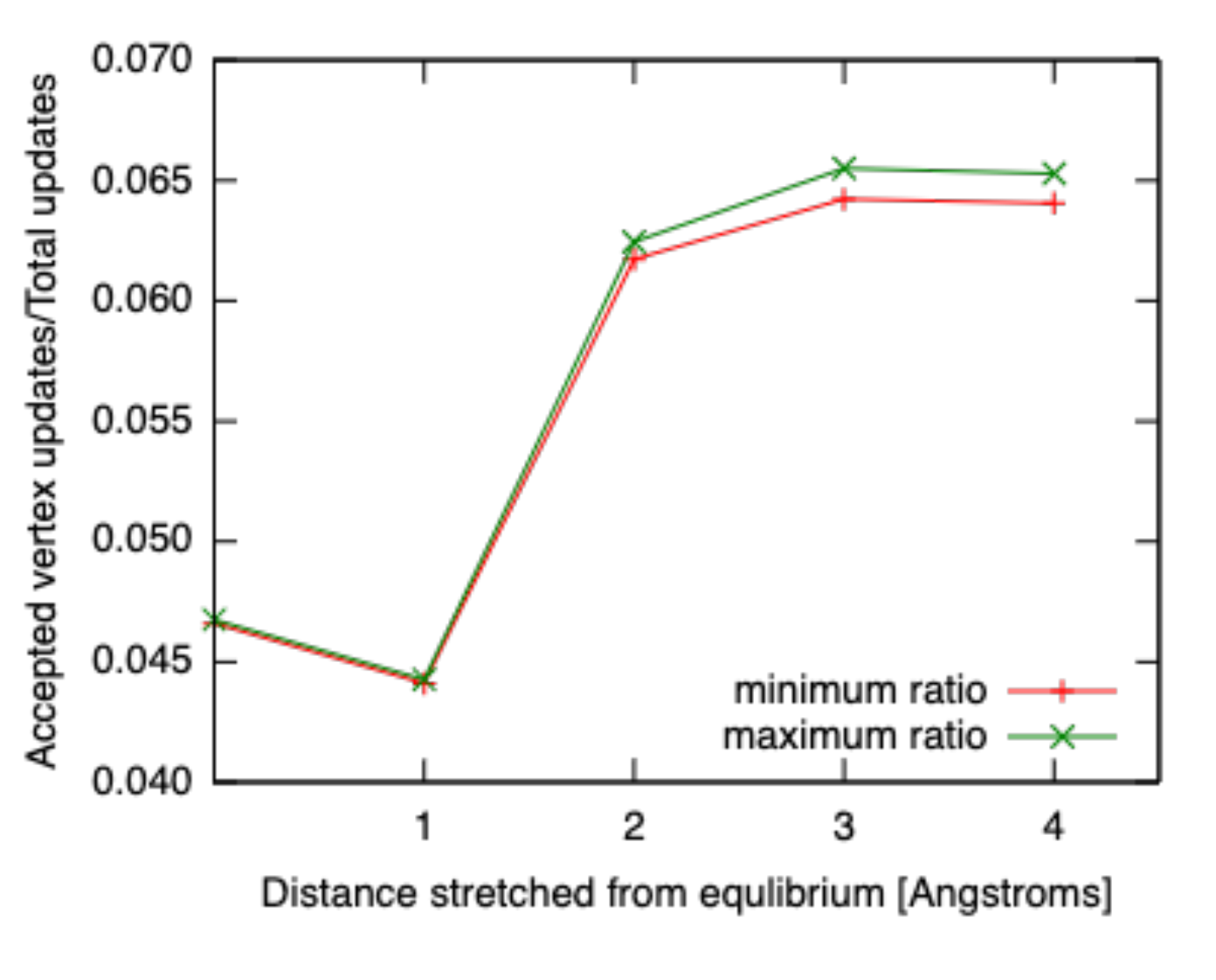}
     \end{center}
  \caption{Update acceptance rates when stretching a water trimer in the cc-pVDZ basis, as a function of stretching distance. Left panel:  acceptance rate of all updates (changing $\tau\alpha\beta\lambda\mu\nu\sigma$). Middle panel: Green's function index update acceptance rate (changing $\lambda\mu\nu\sigma$). Right panel: vertex index update acceptance rate (changing $\alpha\beta$).}
  \label{fig:water_trimer_combined}
\end{figure*}

\subsection{Update efficiency: Stretched Water Trimer}
Next, we show the behavior of the importance sampling procedure as a molecule is stretched. As an example, we study a water trimer in the cc-pVDZ basis.  Subsequently, we stretch the water molecules by taking the individual water molecules from the midpoint and placing them 1 {\AA} further from the midpoint. For details concerning the geometries used, see the Supplementary Info.
 The number of Monte Carlo steps is $10^8$, the normalization subspace is chosen as $n_t^v=8$. 
In Fig.~\ref{fig:water_trimer_combined}, we show how the number of accepted orbital, vertex, and Green's function updates changes with stretching of intermolecular distance.

The flattening of the acceptance ratios at large distance indicates that the sampling is almost entirely performed within a monomer and that the algorithm is capable of avoiding sampling contributions that tend to zero due to the separation distance.

\subsection{Update efficiency: Glycine Peptide Chain}
\begin{figure}
     \includegraphics[width=\columnwidth]{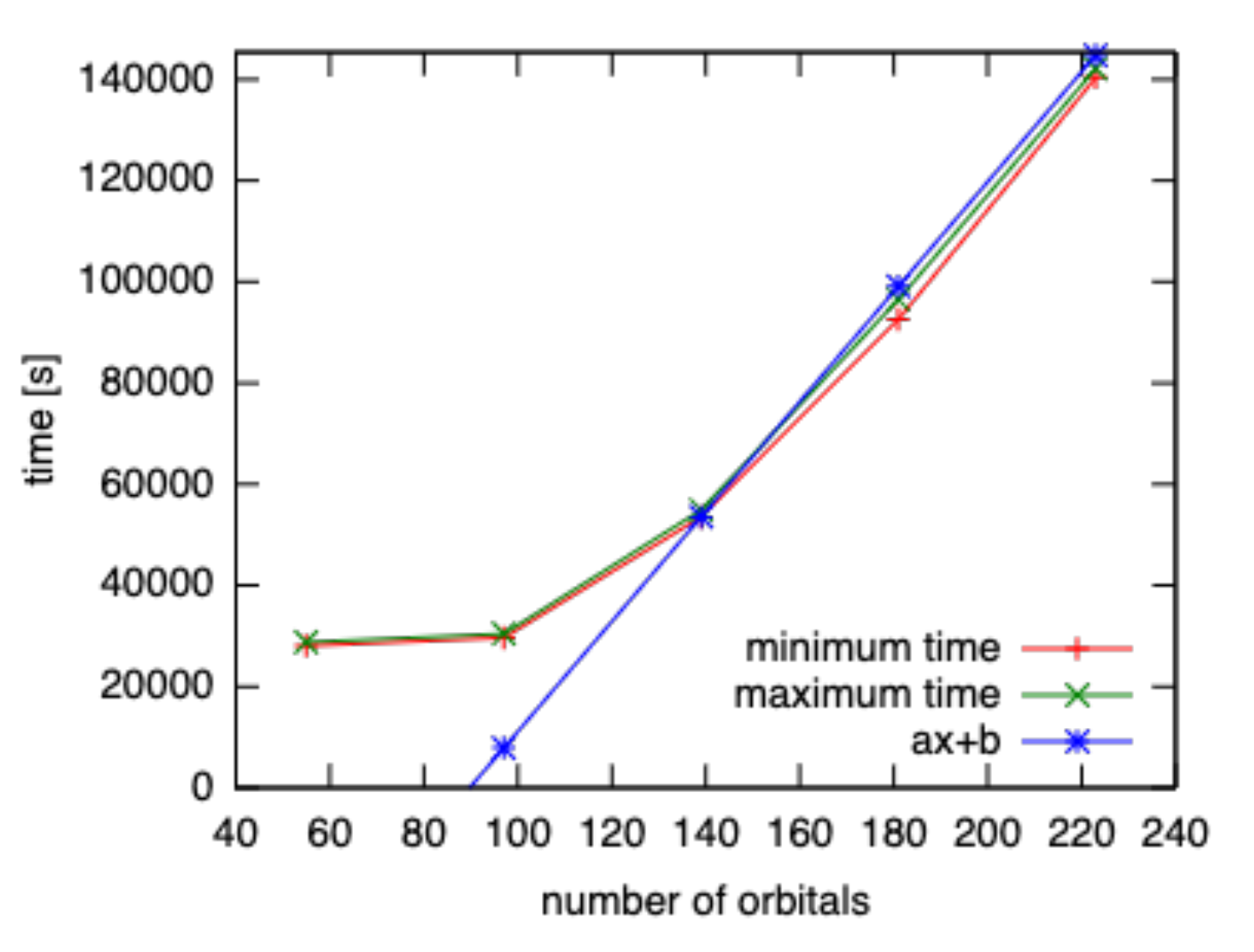}
  \caption{Timing in seconds (y-axis) for  peptide chains containing Glycine residues in the 6-31G basis. The number of orbitals present in the chain is plotted on the x-axis. We also display a linear curve approximating the data for larger number of orbitals.}
  \label{fig:GlycineTimes}
\end{figure}

Finally, we consider the efficiency of the Monte Carlo updates as the size of the system, or the number of orbitals, increases. For this we studied a series of peptide chains containing Glycine residues in the 6-31G basis.  We studied peptides ranging from 1 to 5 residues. For details concerning the geometries used, see the Supplementary Info. The number of Monte Carlo sampling steps is $10^7$, and a normalization subspace of $n_t^v=15$ is chosen. From Fig.~\ref{fig:GlycineTimes}, we see that in the limit of larger number of glycine monomers, the timing of the calculations is consistent with an almost linear scaling as a function of the number of orbitals. 
Note that the cost of a deterministic calculation would be expected to grow like $\mathcal{O}(n_{\tau}n^5)$, where $n$ is the number orbitals.
The almost linear increase of the simulation time is  expected  in an efficient stochastic algorithm, as it indicates that the sampling is mostly performed within the monomers and there is only a modest time increase due to the sampling of inter-monomer contributions. 

\section{Conclusions}\label{sec:conclusion}
In conclusion, we have presented a scheme to sample the Luttinger-Ward functional at second order and measure the self-energy, as well as discussed in detail the error analysis needed for reliable control of error propagation in self-consistent stochastic methods. All two-electron integrals and Green's functions in the Luttinger-Ward functional have been decomposed and compressed as far as possible. A Metropolis importance sampling procedure was then suggested to sample the decomposed space.  To normalize the importance sampling method, we impose a truncation of the decomposition to deterministically solve a subset of the decomposed Luttinger-Ward functional.   

To calibrate the stochastic second-order Green's function method, we separated our analysis into two parts. First, we established that the jackknife resampling method after one iteration of the self-consistent scheme correctly accounts for the non-linear error propagation in the energy equations. We showed this at the example of a toy system consisting of a chain of 10 hydrogens in a minimal basis.  The calculations show that the error bars of the one-body and two-body energies behave as expected, in contrast to naive estimates that lead to incorrect error estimates.  We then validated the convergence of the total energy in a $4\times4$ plaquette of Hydrogens in a minimal basis to the correct deterministic value.

Finally, we explored the behavior of the importance sampling procedure in the water trimer and the Glycine peptide chain.
Here, as expected when Green's functions and two-electron integrals become more sparse, the Metropolis algorithm showed a leveling off in the number of samples required.
We further analyzed the time of simulation for a Glycine peptide chain of 1-5 residues.  We observed that in the limit of larger number of residues there is a linear relationship between the number of orbitals and the time of simulation.

Our results show that this robust, efficient, and fully self-consistent method avoids the bias typically introduced by solving non-linear equations with stochastic results. When sufficiently optimized, our method should be able to treat larger systems that may not be accessible by the deteriministic GF2 algorithm.

\subsection{Acknowledgements}
B.W. and D.Z. acknowledge NSF grant CHE-1453894. 
E.G. was sponsored by the Simons foundation via the Simons Collaboration on the Many-Electron Problem.
\bibliographystyle{apsrev4-1}

\end{document}